
\magnification=1200
\hsize=15truecm
\vsize=23truecm
\baselineskip 18 truept
\voffset=-0.5truecm
\parindent=0cm
\overfullrule=0pt
\font\titolo=cmbx10 scaled\magstep2

\def\Ai{\hbox{\hbox{${\cal A}$}}\kern-1.9mm{\hbox{${/}$}}}
\def\Vi{\hbox{\hbox{${\cal V}$}}\kern-1.9mm{\hbox{${/}$}}}
\def\Di{\hbox{\hbox{${\cal D}$}}\kern-1.9mm{\hbox{${/}$}}}
\def\lam{\hbox{\hbox{${\lambda}$}}\kern-1.6mm{\hbox{${/}$}}}
\def\D{\hbox{\hbox{${D}$}}\kern-1.9mm{\hbox{${/}$}}}
\def\A{\hbox{\hbox{${A}$}}\kern-1.8mm{\hbox{${/}$}}}
\def\V{\hbox{\hbox{${V}$}}\kern-1.9mm{\hbox{${/}$}}}
\def\parz{\hbox{\hbox{${\partial}$}}\kern-1.7mm{\hbox{${/}$}}}
\def\B{\hbox{\hbox{${B}$}}\kern-1.7mm{\hbox{${/}$}}}
\def\R{\hbox{\hbox{${R}$}}\kern-1.7mm{\hbox{${/}$}}}
\def\si{\hbox{\hbox{${\xi}$}}\kern-1.7mm{\hbox{${/}$}}}

\null
\vskip 1.5truecm
\rightline{DFPD/95/TH/44}
\rightline{August 1995}
\vskip 1truecm
\centerline{\titolo Seven--Superform Gauge Fields in N=1, D=10}

\centerline{\titolo Supergravity and Duality$^*$}

\vskip 0.5truecm
\centerline{K. Lechner and M.Tonin}

\centerline{\it Dipartimento di Fisica, Universit\`a di Padova}
\centerline{\it and}
\centerline{\it Istituto Nazionale di Fisica Nucleare, Sezione di Padova}
\centerline{\it Italy}

\vskip 2truecm
\midinsert
\baselineskip 15truept
\centerline{\bf Abstract}

\vskip 0.3truecm

\qquad

We present a formulation of $N=1,D=10$ Supergravity--Super--Maxwell theory
in superspace in which the graviphoton can be described by a 2--form $B_2$
or a 6--form $B_6$, the photon by a 1--form $A_1$ or a 7--form $A_7$ and
the dilaton by a scalar $\varphi$ or an 8--form $\varphi_8$, the
supercurvatures of these fields being related by duality. Duality
interchanges Bianchi identities and equations of motion for each of the
three couples of fields. This construction envisages the reformulation of
$D=10$ Supergravity, involving 7--forms as gauge fields, conjectured by
Schwarz and Sen, which, upon toroidal compactification to four dimensions,
gives the manifestly $SL(2,R)_S$ invariant form of the heterotic string
effective action.

\endinsert

\vskip 2truecm
$^*$ Supported in part by M.P.I.. This work is carried out in the framework
of the European Community Programme ``Gauge Theories, Applied Supersymmetry
and Quantum Gravity" with a financial contribution under contract
SC1--CT92--D789.

\vfill\eject

\qquad
Recently Schwarz and Sen [1] got a manifestly general coordinate invariant
four dimensional heterotic string effective
action which is also manifestly invariant under the
$SL(2,R)_S$ duality.
This action follows quite naturally from dimensional reduction of the dual
version of $D=10$ Supergravity. However, manifest $SL(2,R)_S$ invariance
could not be
achieved when trying to include the Super--Yang--Mills (SYM) sector
in the ten dimensional model. For that, it would
be necessary to replace half of the four dimensional scalar and vector
fields arising from the abelian, Cartan subalgebra valued, ten dimensional
gauge fields with their duals. Schwarz and Sen suggested that these dual
fields should arise, via dimensional reduction, from 7--forms in $D=10$.
This led them to conjecture that a new formulation of
$N=1,D=10$ Supergravity--Super--Maxwell (SUGRA--MAX) should
exist where half of the abelian gauge fields are replaced by abelian
7--forms.

\qquad
This letter provides the main ingredients for this new formulation. We shall
work in the framework of the geometric superspace approach (see ref. [2]
and references therein). In this approach one introduces supervielbeins,
Lorentz and gauge superconnections and, in case, $p$--superforms, imposes
suitable constraints on their curvatures and solves the Bianchi identities
(B.I.) under these constraints to get the equations of motion and the
supersymmetry transformation laws for the physical fields.

\qquad
A recent discussion of $N=1, D=10$ Supergravity models in this framework
has been made in [3]. In this letter we shall follow the notations of that
paper. In particular, superspace is described locally by the
supercoordinates $Z^M = (X^m, \theta^\mu)$ where $X^m$
$(m=0,\cdots,9)$
are space--time coordinates and $\theta^\mu$ $(\mu =1,\cdots,16)$
are Grassmann variables. Latin
and Greek letters denote respectively vector--like and spinor--like
indices and Capital letters both kind of indices. Letters from the
beginning of the alphabet are kept for the (co) tangent superspace.
Given a local frame specified by the one superforms $E^A (Z)$ and a
$p$--superform $\Psi_p$

$$
\Psi_p = {1 \over p!} E^{A_1} \cdots E^{A_p} \Psi_{A_p \cdots A_1}
$$

it will be useful to call $(q, p-q)$ sector of $\Psi_p$, denoted by
$\Psi_{q, p-q}$, the component of $\Psi_p$ proportional to $q$ vector--like
supervielbeins $E^a$ and $p-q$ spinor--like supervielbeins $E^\alpha$.
Moreover,
$d$ is the superspace differential and $D$ is the Lorentz--covariant one.

\qquad
In the standard superspace formulation of $N=1$, $D=10$ SUGRA--SYM one
introduces the supervielbeins $E^A (Z)$, the Lorentz valued superconnection
$\Omega_A{}^B (Z)$ the Lie algebra valued gauge superconnection $A_1(Z)$ and
the 2--superform $B_2 (Z)$. Their curvatures are respectively the torsion
$T^A=DE^A$, the Lorentz curvature $R_A{}^B$, the gauge curvature $F_2$ and the
$B_2$--curvature $H_3 = dB_2 + \Omega_3$.
The 3--superform $\Omega_3$ depends on the model. For instance
it vanishes in pure supergravity and it is proportional to the Chern--Simons
3--superform associated to $A_1$ in the minimally coupled SUGRA--SYM [4].
The theory is set on shell by imposing suitable constraints on the
lower dimension sectors of torsion, gauge curvature  and $B_2$--curvature
[5].

\qquad
In the dual formulation [6] $B_2$ is replaced by a 6--superform $B_6$.
However in this case the theory can not be put on-shell by imposing
suitable constraints on the $B_6$--curvature, $H_7=dB_6$, in addition
to the standard torsion and gauge curvature constraints [7].
For that one has to  fix a superfield, which belongs to the
120 irreducible representation (irrep.) of $SO(1,9)$,
which can not be determined by solving the B.I., see [8,9].
For a recent discussion of the dual formulation see [9].

\qquad
A more symmetric situation arises in the approach of ref. [3]. Here the
superforms $B_2$ and $B_6$ are not even introduced from the beginning and
the theory is set on shell by adding to the usual torsion and gauge
curvature constraints a further constraint on the $(0,2)$ sector
of the Lorentz curvature. It has been shown in [3] that under these
constraints one can define a 3--superform $H_3$ and a 7--superform $H_7$
that satisfy suitable identities such that $H_3$ ($H_7$) can be
considered locally as the curvature of a 2--superform $B_2$
(6--superform $B_6$). (The asymmetry between the two possibilities,
however, remains since $H_7$ is always closed whereas $H_3$ is closed only
in pure supergravity).

\qquad
In this letter we will show that a situation similar to the one described
in [3] for the $B$--sector arises in the gauge sector and in the dilaton
sector. We shall discuss the minimally coupled SUGRA--MAX theory
with abelian gauge group $U(1)^{2n}$ (in the case of the heterotic string
effective action one has to set  $n=8$). The gauge curvatures are

$$
\vec F_2 = d \vec A_1 \eqno(1)
$$

with the B.I.

$$
d \vec F_2 =0. \eqno(2)
$$

Here $\vec A_1 \equiv (A_1^{(1)},\cdots, A_1^{(2n)})$ denotes the set of
Maxwell one--superforms.
In this case the identity for $H_3$ is

$$
d H_3 = \gamma  \vec F_2 \cdot \vec F_2 \eqno(3)
$$

so that locally

$$
H_3 = dB_2 + \gamma \vec A_1 \cdot  \vec F_2, \eqno(4)
$$

and the identity for $H_7$ is

$$
dH_7 =0 \eqno(5)
$$

so that locally

$$
H_7 = dB_6. \eqno(6)
$$

\qquad
We shall show that under the constraints of ref. [3] it is possible to
reconstruct $2n$  eight--superforms $\vec F_8$, dual to $\vec F_2$,
$\vec F_8 \equiv (F_8^{(1)},
\cdots, F_8^{(2n)})$,
that satisfy the identities

$$
d \vec F_8 = \vec F_2 H_7, \eqno(7)
$$

so that locally there exist $2n$ 7--superforms $\vec A_7$ $(Z)$ such that

$$
\vec F_8 = d \vec A_7 + \vec F_2 B_6. \eqno(8)
$$

\qquad
In a formulation where $\vec A_1$ are the relevant gauge fields the sector
(3,0) of eq. (2) describes the B.I. for their curvatures and the sector
(9,0) of eq. (7) provides their field equations.
However, according to eq. (8), we propose
other formulations in which some
of the one--superforms $\vec A_1$ can be replaced by
the corresponding 7--superforms $\vec A_7$,
in agreement with the conjecture of ref. [1]. In this case (7) becomes a
B.I. and (2) has to be read as equation of motion for $\vec A_7$.

\qquad
In addition we shall show that it is possible to reconstruct a
9--superform $V_9$, dual to the curvature $V_1$ of the dilaton
$\varphi$, $V_1=d\varphi$, that satisfies the identity

$$
dV_9 = H_3 H_7 + \gamma  \vec F_2 \cdot \vec F_8, \eqno(9)
$$

in such a way that the highest sector (10, 0) of eq. (9) is just the field
equation for $\varphi$ (a 0--superform). Again, (9) implies
locally the existence of an 8--superform $\varphi_8$ such that

$$
V_9 = d \varphi_8 - B_2H_7 + \gamma \vec A_1 \cdot \vec F_8, \eqno(10)
$$

or, alternatively, of an 8--superform $\hat\varphi_8$ such that
$$
V_9 = d \hat\varphi_8 + H_3B_6 + \gamma \vec F_2 \cdot \vec A_7. \eqno(11)
$$
If (9) is regarded as the B.I. for $(\varphi_8,\hat\varphi_8)$ its
equation of motion becomes just

$$
dV_1=0. \eqno(12)
$$

\qquad
To prove these results we shall work for simplicity in the
SUGRA--MAX theory with only one gauge multiplet, the
generalization to the case with $2n$ gauge multiplets  being straightforward.
The relevant B.I. to be considered are the torsion B.I.

$$
DT^A = E^B R_B{}^A \eqno(13)
$$

and the gauge B.I.

$$
d F_2=0. \eqno(14)
$$

The torsion and curvature constraints, chosen in [3], are

$$
T_{\alpha\beta} {}^a = 2 (\Gamma^a)_{\alpha\beta}, \qquad
T_{a \alpha}{}^b =0=
T_{\alpha a}{}^b \eqno(15)
$$
$$
F_{\alpha\beta} =0 \eqno(16)
$$
$$
R_{\alpha\beta ab} = (\Gamma_{[a} \Gamma^{cde}
\Gamma_{b]})_{\alpha\beta} J_{cde} \footnote{*}
{This constraint differs from the
one in [3] by a shift of the Lorentz connection.}.  \eqno(17)
$$
\par
Notice that, apart from (13)--(17), we do not impose constraints or B.I
on any other superform. All our results will be obtained
by demanding the closure of the SUSY--algebra on (13)--(17)
\par
We defined $\Gamma^{a_1 \cdots a_k} \equiv \Gamma^{[a_1} \Gamma^{a_2}
\cdots \Gamma^{a_k]}$ and $(\Gamma^a)_{\alpha\beta},
(\Gamma^a)^{\alpha\beta}$ are Weyl matrices in $D=10$.
The current $J^{cde}$, a 120 irrep. of $SO(1,9)$, is a local function of
the relevant superfields and (13) demands
that in the covariant spinorial derivative of $e^{2 \varphi} J_{cde}$,
the highest irrep., i.e. the 1200, is absent.
The explicit  expression of $J_{cde}$ depends on the model considered.
In the present case it is given by

$$
J_{cde} = -{\gamma \over 12} (\Gamma_{cde})_{\alpha\beta}
\chi^\alpha \chi^\beta \equiv - {\gamma \over 12}\chi_{cde}
\eqno(18)
$$
and it can be verified that it satisfies the just mentioned condition.
Here $\chi^\alpha$ is the gravitino superfield which is present in the
(1,1) sector of the gauge curvature

$$
F_{a \alpha} = 2 (\Gamma_a)_{\alpha\beta} \chi^\beta \eqno(19)
$$

as a consequence of the constraint (16) and the B.I. (14).
$\gamma$ is a coupling constant which vanishes in pure Supergravity.
{}From the
B.I. (13) one has also

$$
T_{\alpha\beta}{}^\gamma = 2 \delta^\gamma_{(\alpha} V_{\beta)}  -
(\Gamma^a)_{\alpha\beta} (\Gamma_a)^{\gamma\delta} V_\delta \eqno(20)
$$

$$
T_{a \alpha}{}^\beta = {1 \over 4} (\Gamma^{bc})_\alpha{}^\beta T_{abc} -
2 \gamma (\Gamma_a)_{\alpha\gamma} \chi^\gamma \chi^\beta \eqno(21)
$$

$$
R_{a \alpha bc} = 2(\Gamma_a)_{\alpha\beta} T_{bc}{}^\beta + 6 \gamma
(\Gamma_{[a})_{\alpha\beta} F_{bc]} \chi^\beta \eqno(22)
$$

where
$V_\alpha \equiv D_\alpha \varphi$
is the gravitello superfield, $T_{abc}=T_{ab}{}^d \eta_{dc}$ belongs
to the 120 irrep. and $T_{bc}{}^\beta$ is the gravitino field strength.
Moreover, the eqs. (13)--(17) imply the following relations

$$
D_\alpha V_\beta = - \Gamma^a_{\alpha \beta} D_a \varphi + V_\alpha V_\beta +
{1\over 12} (\Gamma_{abc})_{\alpha\beta} \left(T^{abc} + {\gamma \over 2}
\chi^{abc}\right) \eqno(23a)
$$

$$
D_\alpha T_{abc} = (\Gamma_{[a})_{\alpha\beta}
\left(-6 T_{bc]}{}^\beta - 12 \gamma F_{bc]}
\chi^\beta\right) \eqno(23b)
$$

$$
D_\alpha \chi^\beta = {1\over 4} (\Gamma_{ab})_\alpha{}^\beta F^{ab} +
T_{\gamma\alpha}{}^\beta \chi^\gamma \eqno(23c)
$$

$$
D_\alpha F_{ab} = 4(\Gamma_{[a})_{\alpha\beta} D_{b]} \chi^\beta+
(\Gamma^{cd}{}_{[a})_{\alpha\beta} T_{b] cd} \chi^\beta \eqno(23d)
$$

$$
D_{[a} T_{bcd]} + {3\over 2} T^f{}_{[ab} T_{cd]f} = {3 \gamma \over 2}
F_{[ab} F_{cd]}
\eqno(24)
$$

$$
D_c T^c {}_{ab} = 2 D_c \varphi T^c {}_{ab} + 2 T_{ab}{}^\alpha
V_\alpha + \gamma\left(
4 F_{ab} \chi^\alpha V_\alpha
-4\chi_\alpha (\Gamma_{[a})^{\alpha\beta} D_{b]} \chi_\beta + T_{[a}
{}^{cd} \chi_{b]cd}\right),
\eqno(25)
$$

together with the field equations:

$$
(\Gamma^a)^{\alpha \beta}  D_a V_\beta = 2(\Gamma^a)^{\alpha \beta}
D_a \varphi  V_\beta - {1 \over 12}
(\Gamma^{abc})^{\alpha \beta} T_{abc} V_\beta +  \gamma \left(2\chi^\alpha
\chi^\beta V_\beta - {1 \over 2} (\Gamma^{ab})_\beta{}^\alpha F_{ab}
\chi^\beta\right) \eqno(26)
$$

$$
D^a D_a \varphi = 2 D^a \varphi D_a \varphi - {1\over 12} T^{abc} T_{abc}
$$
$$
+ \gamma\left(
{1 \over 12} T^{abc} \chi_{abc} - {1\over 96} \chi^{abc}
V_{abc} - {1\over 4} F^{ab} F_{ab} + {1 \over 4}
(\Gamma^{ab})_\alpha {}^\beta V_\beta \chi^\alpha F_{ab}\right) \eqno(27)
$$

$$
(\Gamma^b)_{\alpha\beta} T_{ba}{}^\beta = D_a V_\alpha + {1\over 4}
(\Gamma^{bc})_\alpha{}^\beta V_\beta T_{a bc} - \gamma
(\Gamma^b)_{\alpha\beta} F_{ba} \chi^\beta \eqno(28)
$$

$$
R_{(ab)} = 2 D_{(a} D_{b)} \varphi + \gamma\left(2 \chi^\alpha
(\Gamma_{(a})_{\alpha\beta} D_{b)} \chi^\beta - F_a{}^c
F_{bc} + {1\over 2} \chi_{(a}{}^{cd} T_{b)cd}\right) \eqno(29)
$$

$$
(\Gamma^a)_{\alpha\beta} D_a \chi^\beta
=\left((\Gamma^b)_{\alpha\beta} D_b \varphi + V_\alpha V_\beta -
{1\over 6} (\Gamma^{abc})_{\alpha\beta} T_{abc} \right) \chi^\beta  -
{1\over 4} (\Gamma^{ab})_\alpha{}^\beta F_{ab} V_\beta \eqno(30)
$$

$$
D^b F_{ba} = 2 D^b \varphi F_{ba} + 2 V_\beta D_a \chi^\beta + T_{abc} F^{bc} -
{1\over 2} T_{abc} (\Gamma^{bc})_\beta{}^\gamma V_\gamma \chi^\beta.
\eqno(31)
$$

Here we defined

$$
V^{abc} = (\Gamma^{abc})^{\alpha\beta} V_\alpha V_\beta.
$$

Eqs. (26) -- (31) are the field equations for the
gravitello, the dilaton,
the gravitino, the graviton, the gaugino and the gauge boson respectively.

\qquad
According to ref. [3], using the relations
above, one can define a 3--superform $H_3$ with components

$$\eqalign{
H_{abc} & = T_{abc} \cr
H_{a \alpha\beta} & = 2 (\Gamma_a)_{\alpha \beta}, \cr} \eqno(32)
$$

and all others vanishing,
such that (see for instance (24))

$$
d H_3 = \gamma F_2 F_2, \eqno(33)
$$

and a 7--superform $H_7$ with components

$$
H_{a_1 \cdots a_7} = {1 \over 3!} e^{-2 \varphi} \varepsilon_{a_1 \cdots
a_7}{}^{b_1 b_2 b_3}  (H_{b_1 b_2 b_3} - V_{b_1 b_2 b_3} - \gamma \chi_{b_1
b_2 b_3}) \eqno(34a)
$$

$$
H_{\alpha a_1 \cdots a_6} = - 2 e^{-2 \varphi} (\Gamma_{a_1 \cdots
a_6})_\alpha{}^\beta V_\beta \eqno(34b)
$$

$$
H_{\alpha \beta a_1 \cdots a_5} = -2 e^{-2 \varphi} (\Gamma_{a_1 \cdots
a_5})_{\alpha\beta} \eqno(34c)
$$

and all others vanishing, such that

$$
dH_7 =0. \eqno(35)
$$

It follows from (33) and (35) that one can write locally

$$
H_3 = d B_2 + \gamma A_1 F_2 \eqno(36)
$$

and

$$
H_7 = dB_6. \eqno(37)
$$

Of course $B_2$ and $B_6$ are not independent superforms since their
curvatures are related through eq. (34a) and can be considered
as dual one to the other. If one works with $B_2$ eq. (33) is the B.I. and
eq. (35) provides the field equation of $B_2$ and, viceversa, working with
$B_6$ eq. (35) is the B.I. and eq. (33) contains the field equation of
$B_6$.

\qquad
Going to the gauge sector, let us notice at first that the gauge boson
field equation (31), using the gluino, gravitello and gravitino equations
of motion, can be rewritten as

$$
e^{2\varphi}D^b \left(e^{-2 \varphi} \tilde F_{ba}\right) -
{1\over 2}
T_{abc} \tilde F^{bc} -(\Gamma_a{}^{bc})_{\alpha\beta}
T_{bc}{}^\beta \chi^\alpha = {1 \over 2} (H_{abc} -
V_{abc} - \gamma \chi_{abc}) F^{bc} \eqno(38)
$$

where

$$
\tilde F^{ab} \equiv F^{ab} + 2 (\Gamma^{ab})_\alpha{}^\beta V_\beta
\chi^\alpha.
$$

Now let us define the 8--superform $F_8$ through

$$
F_{a_1 \cdots a_8} = {1 \over 2} e^{-2\varphi}
\varepsilon_{a_1 \cdots a_8}{}^{b_1 b_2} \tilde F_{b_1 b_2}  \eqno(39a)
$$

$$
F_{\alpha a_1 \cdots a_7}= -2 e^{-2 \varphi} (\Gamma_{a_1 \cdots a_7})_
{\alpha\beta} \chi^\beta, \eqno(39b)
$$
and all other components vanishing.
A lengthy but straightforward calculation shows that $F_8$ satisfies the
identity

$$
dF_8 = F_2 H_7, \eqno(40)
$$

so that locally there exists a 7--superform $A_7$ such that

$$
F_8 = dA_7 + F_2 B_6. \eqno(41)
$$

\qquad
The simplest way to prove (40) is to define
$Y_9 \equiv dF_8 - F_2 H_7$. $Y_9$ vanishes trivially in the sectors
(5,4), (4,5),$\cdots$,(0,9).
A relatively easy calculation shows that it vanishes also in the sectors
(6,3) and (7,2). Then, looking at the identity $dY_9 =0$ in the sectors
(7,3) and (8,2), one can see immediately that $Y_9$  vanishes also in the
sectors (8,1) and (9,0). However, it is instructive to verify directly
that (the dual of) eq. (40) in the sector (9,0) is precisely eq. (38).

\qquad
As for the dilaton sector, the use
of the equation of motion of the gravitino allows to rewrite the dilaton field
equation (27) in the form

$$
\eqalign
{e^{2\varphi}
D_a \left(e^{-2\varphi} D^a \varphi\right) - {1\over 2}
(\Gamma^{ab})_\alpha{}^\beta T_{ab}{}^\alpha V_\beta = &- {1\over 12}
 H_{abc} \left(H^{abc} - V^{abc} - \gamma \chi^{abc} \right)\cr
&- {\gamma \over 4} F_{ab} \tilde F^{ab}.\cr
}
\eqno(42)
$$

Then we can define the 9--superform $V_9$  with components

$$
V_{a_1 \cdots a_9} = 2 e^{-2 \varphi} \varepsilon_{a_1 \cdots a_9}{}^b
D_b \varphi \eqno(43a)
$$

$$
V_{\alpha a_1 \cdots a_8}= -2 e^{-2 \varphi} (\Gamma_{a_1 \cdots
a_8})_\alpha {}^\beta V_\beta
\eqno(43 b)
$$
and all others vanishing,
and verify, as before, that $V_9$ satisfies the identity

$$
dV_9 = H_3 H_7 + \gamma F_2 F_8. \eqno(44)
$$

The (dual of) eq. (44) in the sector (10,0) is just the dilaton field
equation (42). It follows from (44) that locally there exists an
8--superform $\varphi_8$ such that

$$
V_9 = d \varphi_8 - B_2 H_7 + \gamma A_1 F_8, \eqno(45)
$$

or, alternatively, $\hat\varphi_8$ such that

$$
V_9 = d \hat\varphi_8 +H_3B_6 + \gamma F_2A_7. \eqno(46)
$$

\qquad
In models with $2n$ Super--Maxwell multiplets, some of the field equations
and supersymmetry transformations become slightly more complicated
(see, for instance [10] where a set of constraints similar to ours has been
used). However, it is straightforward to extend our results to this case
to get again eqs. (7) -- (11) where the $F^{(r)}_8$ are still given by eq.
(39) for  each $r$ $(r = 1, \cdots, 2n)$.

\qquad
As $B_2$ and $B_6$ in the $B$ sector, also $A_1$ and $A_7$ in the gauge
sector are not independent superforms and must be considered as dual one
to each other. The same happens for $\varphi$ and $\varphi_8$ in the
dilaton sector.
Then one can foresee different formulations of $N=1, D=10$
SUGRA--MAX models (with gauge group $U(1)^{2n}$)
where $B_2$ or $B_6$, $A_1^{(r)}$ or $A_7^{(r)}$ for each
$r$, $\varphi$ or $(\varphi_8,\hat\varphi_8)$ are
considered as the fundamentals fields. If one
chooses $A_1^{(r)}$ ($A_7^{(r)}$)  eq. (2) (eq. (7)) is the B.I.
and eq. (7) (eq. (2))
provides the relevant equation of motion. Similarly for $\varphi$
($\varphi_8,\hat\varphi_8$) eq. (12) (eq. (9))
is the B.I. and eq. (9) (eq. (12))
provides the relevant equation of motion.

\qquad
However, notice that not all combinations of $B_2$ or $B_6$, $A_1$ or
$A_7$, $\varphi$ or $\varphi_8$ are allowed. Indeed, taking a look on
eq. (8) one sees that $A_7$
is compatible with $B_6$ but not with $B_2$. As for
$(\varphi_8,\hat\varphi_8)$, it is convenient to rescale $B_6,A_7,(\varphi_8
,\hat\varphi_8)$ as well as $H_7,F_8,V_9$ by the factor $e^{2\varphi}$
in order to get rid of any dependence on $\varphi$ in eqs. (45),(46) and
(41). Then one can see that $\varphi_8$ is compatible only with $B_2,A_1$
while $\hat\varphi_8$ is compatible only with $B_6,A_7$.

\qquad
Among the allowed formulations that in terms of ($\varphi,A_1,B_2$) is
the standard formulation and that in terms of ($\varphi,A_1,B_6$) is
the "old" dual one. The new formulation advocated in ref. [1] corresponds to
the one formulated in terms of ($\varphi,B_6,A_1^{(r)},A_7^{(r+n)},
r=1,\cdots,n$).
The previous remark about the compatibility of $A_7$ with $B_6$ but not
with $B_2$ is in agreement with the result of [1], namely that the
ten-dimensional supergravity
version which leads to the manifestly $SL(2,R)_S$
invariant action, after toroidal compactification to four dimensions,
is the one which involves $B_6$.

\qquad
We must point out a complication that arises when working with $A_7$
instead of $A_1$:
when in eq. (41) $F_2$ is removed in favour of
$F_8$, eq. (41) cannot be inverted in a closed form to get $F_8$ in
terms of $A_7$. The best one can do is to express $F_8$ in terms of $A_7$
as an iterative series.
A similar complication arises in the dilaton sector, eqs. (45),(46).
Nevertheless this
situation is not new; it was also met for instance
in the equation for $H_{abc}$ of
the anomaly free models with Lorentz Chern--Simons
coupling, see [11].

\qquad
Another feature of eq. (41) has to be pointed out: in order to get
an $F_8$ invariant
under the gauge transformation of $B_6$,
$$
\delta B_6 = d \Lambda_5, \eqno(47)
$$

$A_7$ too has to transform as

$$
\delta A_7 = -F_2 \Lambda_5. \eqno(48)
$$

This is analogous to the fact that $B_2$ in the usual formulation,
transforms under gauge (and Lorentz) transformations as a consequence of
the gauge (and Lorentz) Chern--Simons forms in the definition of $H_3$.
However, here (41) is invariant under (47) and (48) only if eq. (2) is
satisfied, which now has to be interpreted as the equation of
motion of $A_7$, so that invariance of eq. (41) holds only on--shell.

\qquad
Let us comment finally on what our results become in two more general
situations. If one considers a non abelian gauge group eq. (40) does
no longer hold since in this case the gluinos are minimally coupled
to the gluons and (40) becomes now $DF_8=F_2H_7+j_9$ where $j_9$
is a current nine-superform which is Lie-algebra-valued together with
$F_2$ and $F_8$. In this case it is not possible to replace $A_1$ by
$A_7$, nor to replace $\varphi$ by ($\varphi_8,\hat\varphi_8$).
Nevertheless eq. (44) holds again upon tracing the $F$-terms
on its right hand side. The r.h.s. of (44) is in this case
a closed form, thanks to the identity $tr(j_9F_2)=0$, but not an exact one
as in the abelian case.

\qquad
On the contrary, if one keeps the gauge group abelian but introduces
the Lorentz-Chern-Simons three form in the definition of $H_3$
(which amounts to a modification of $J_{cde}$ in (18)) it can be seen
that (40) holds again upon substituting the term $\gamma\chi_{b_1b_2b_3}$
in (34a) -- and in due places --
with $-12 J_{b_1b_2b_3}$. In fact, the sectors $(6,3)$ and
$(7,2)$ of $Y_9$, defined after eq. (41), are independent on $J_{abc}$ and
its superspace derivatives, and again $dY_9=0$ due to $dH_7=0$.
On the other hand now (44) does no longer hold.
Therefore in this case it is still possible to describe the gauge degrees
of freedom in terms of $A_7$, while the dilaton has to be described
by a scalar.

\qquad
It may also be that our results shed some new light on
the string/fivebrane or string/string duality relations conjectured
recently and in the
past (see ref. [12] and references therein).

\vskip1truecm

{\bf References}
\vskip 0.3truecm
[1] J.H. Schwarz and A. Sen, Nucl. Phys. \underbar{B411}, 35 (1994).

\smallskip
[2] L. Castellani, R. D'Auria and P. Fr\`e: ``Supergravity and
Superstrings: a geometric prospective". Vol. 1--3 (World Scientific,
Singapore, 1991).

\smallskip
[3] A. Candiello and K. Lechner, Nucl. Phys. \underbar{B412}, 479 (1994).

\smallskip
[4] A. Chamseddine, Nucl. Phys. \underbar{B185}, 403 (1981);
G.F. Chaplin and N.S. Manton, Phys. Lett. \underbar{120B}, 105 (1983);
R. D'Auria, P. Fr\`e and A. J. da Silva, Nucl. Phys. \underbar{B196}, 205
(1982);
E. Bergshoeff, M. de Roo, B. de Wit and P. van Nieuwenhuizen Nucl. Phys.
\underbar{B195}, 97, (1982).

\smallskip
[5] B.E.W. Nilsson, Nucl. Phys. \underbar{B188}, 176 (1981);
Goteborg preprint 81--6 (1981);
P. Howe, H. Nicolai and A. van Proeyen, Phys. Lett. \underbar{B112}, 446
(1982);
E. Witten, Nucl. Phys. \underbar{B266}, 245 (1986);
J.J. Atick, A. Dhar and B. Ratra, Phys. Rev. \underbar{D33}, 2824 (1986);
B.E.W. Nilsson and A. Tollsten, Phys. Lett. \underbar{169B} 369 (1986);
B.E.W. Nilsson and R. Kallosh, Phys. Lett. \underbar{167B} 46 (1986).

\smallskip
[6] A. Chamseddine, Phys. Rev. \underbar{B24}, 3065 (1981);
L. Castellani, P. Fr\`e, F. Pilch and P. Van Nieuwenhuizen, Ann. of Phys.
\underbar{146}, 35 (1983);
S.J. Gates and H. Nishino, Phys. Lett. \underbar{B173} 46, (1986); Nucl.
Phys. \underbar{B291}, 205 (1987).

\smallskip
[7] R. D'Auria and P. Fr\`e, Mod. Phys. Lett. \underbar{A3}, 673 (1988).

\smallskip
[8] S. McDowell and M. Rakowski, Nucl. Phys. \underbar{B274} 589 (1986).

\smallskip
[9] M.V. Terentjev, Phys. Lett. \underbar{B313}, 351 (1993);
Phys. Lett. \underbar{B325}, 96 (1994);
H. Nishino, Phys. Lett. \underbar{258B}, 104 (1991).

\smallskip
[10] M. Grisaru, H. Nishino and D. Zanon, Nucl. Phys. \underbar{B314}, 363
(1989).

\smallskip
[11] L. Bonora, P. Pasti and M. Tonin, Phys. Lett. \underbar{B188}, 335
(1987);
L. Bonora, M. Bregola, K. Lechner, P. Pasti and M. Tonin, Nucl. Phys.
\underbar{B296}, 877 (1988) and Int. J. Mod. Phys. \underbar{A5}, 461
(1990);
R. D'Auria and P. Fr\`e, Phys. Lett. \underbar{B200} 63 (1988);
R. D'Auria, P. Fr\`e, M. Raciti and F. Riva, Int. J. Mod. Phys.
\underbar{173} 953 (1988);
M. Raciti, F. Riva and D. Zanon, Phys. Lett. \underbar{B227}, 118 (1989);
K. Lechner, P. Pasti and M. Tonin, Mod. Phys. Lett. \underbar{A2}, 929
(1987);
K. Lechner and P. Pasti, Mod. Phys. Lett. \underbar{A4} 1721 (1989);
I. Pesando, Phys. Lett. \underbar{B272}, 45 (1991).

\smallskip
[12] M. J. Duff, Ramzi R. Khuri and J. X. Lu,
``String Solitons", (1994) HEP-TH/9412184, to appear in Phys. Rep.
\bye